\documentclass[journal]{IEEEtran}
\input epsf
\usepackage{graphicx}
\usepackage{bm}
\usepackage{mathabx}
\usepackage{cite}
\usepackage{tablefootnote}
\usepackage{float}
\usepackage{scalerel}
\usepackage{tikz}
\usetikzlibrary{svg.path}

\definecolor{orcidlogocol}{HTML}{A6CE39}
\tikzset{
  orcidlogo/.pic={
    \fill[orcidlogocol] svg{M256,128c0,70.7-57.3,128-128,128C57.3,256,0,198.7,0,128C0,57.3,57.3,0,128,0C198.7,0,256,57.3,256,128z};
    \fill[white] svg{M86.3,186.2H70.9V79.1h15.4v48.4V186.2z}
                 svg{M108.9,79.1h41.6c39.6,0,57,28.3,57,53.6c0,27.5-21.5,53.6-56.8,53.6h-41.8V79.1z M124.3,172.4h24.5c34.9,0,42.9-26.5,42.9-39.7c0-21.5-13.7-39.7-43.7-39.7h-23.7V172.4z}
                 svg{M88.7,56.8c0,5.5-4.5,10.1-10.1,10.1c-5.6,0-10.1-4.6-10.1-10.1c0-5.6,4.5-10.1,10.1-10.1C84.2,46.7,88.7,51.3,88.7,56.8z};
  }
}

\newcommand\orcidicon[1]{\href{https://orcid.org/#1}{\mbox{\scalerel*{
\begin{tikzpicture}[yscale=-1,transform shape]
\pic{orcidlogo};
\end{tikzpicture}
}{|}}}}

\usepackage{hyperref}
\begin{document}

\title{Flexible Coaxial Ribbon Cable for High-Density Superconducting Microwave Device Arrays}

\author{Jennifer Pearl Smith \orcidicon{0000-0002-0849-5867}, Benjamin A. Mazin  \orcidicon{0000-0003-0526-1114}, Alex B. Walter \orcidicon{0000-0002-4427-4551}, Miguel Daal \orcidicon{0000-0002-1134-2116}, J. I. Bailey, III \orcidicon{0000-0002-4272-263X}, Clinton Bockstiegel, Nicholas Zobrist \orcidicon{0000-0003-3146-7263}, Noah Swimmer \orcidicon{0000-0001-5721-8973}, Sarah Steiger \orcidicon{0000-0002-4787-3285}, Neelay Fruitwala}




\maketitle

\begin{abstract}
Superconducting electronics often require high-density microwave interconnects capable of transporting signals between temperature stages with minimal loss, cross talk, and heat conduction. We report the design and fabrication of superconducting 53 wt\% Nb-47 wt\% Ti (Nb47Ti) FLexible coAXial ribbon cables (FLAX). The ten traces each consist of a ${\diameter}$0.076 mm NbTi inner conductor insulated with PFA (${\diameter}$0.28 mm) and sheathed in a shared 0.025 mm thick Nb47Ti outer conductor. The cable is terminated with G3PO coaxial push-on connectors via stainless steel capillary tubing (${\diameter}$1.6 mm, 0.13 mm thick) soldered to a coplanar wave guide transition board. The 30 cm long cable has 1 dB of loss at 8 GHz with -60 dB nearest-neighbor forward cross talk. The loss is 0.5 dB more than commercially available superconducting coax likely due to impedance mismatches caused by manufacturing imperfections in the cable. The reported cross talk is 30 dB lower than previously developed laminated NbTi-on-Kapton microstrip cables. We estimate the heat load from 1 K to 90 mK to be 20 nW per trace, approximately half the computed load from the smallest commercially available superconducting coax from CryoCoax. 
\end{abstract}
\IEEEoverridecommandlockouts

\IEEEpeerreviewmaketitle

\section{Introduction}
\IEEEPARstart{S}{uperconducting} devices are revolutionizing a wide range of research and technological fields including quantum computing \cite{Barends2014SuperconductingTolerance, Barends2016DigitizedCircuit, Ofek2016ExtendingCircuits, Gu2017MicrowaveCircuits}, nanowire single-photon detectors \cite{Wollman2019KilopixelDetectors}, X-ray microcalorimeters \cite{Ulbricht2015}, submillimeter bolometers \cite{Holland2013SCUBA-2:Telescope}, and Microwave Kinetic Inductance Detectors (MKIDs) \cite{Calvo2016, Mazin2013, Meeker2015, Meeker2018DARKNESS:Astronomy}. These applications require increasingly large superconducting arrays, which present the common technical challenge of transporting microwave signals from the cold device stage to room temperature without losing or corrupting the signal or conducting excess heat to the cold stage. Low thermal conductivity is especially important for detector arrays in the field or in space using adiabatic demagnetization refrigerators (ADRs) which have less cooling power than dilution refrigerators but offer smaller form factors and simpler operation.

Commercially available superconducting coaxial cables are often used below 4 K; however, they are either semi-rigid and cumbersome to use in small cryogenic volumes, have large cross-sections yielding excessive heat loads, or both. Another option is flexible superconducting circuits fabricated using lithography techniques. These laminated cable technologies boast low thermal conductivity and high-density interconnects but lack the length, durability, and signal isolation needed for many applications \cite{Pappas2016High-DensityACTPol, Tuckerman2016FlexibleApplications, Zou2019Low-lossLines, Walter2018, Gupta2019Thin-filmCables}.

\begin{figure}
\includegraphics[width=3.4in]{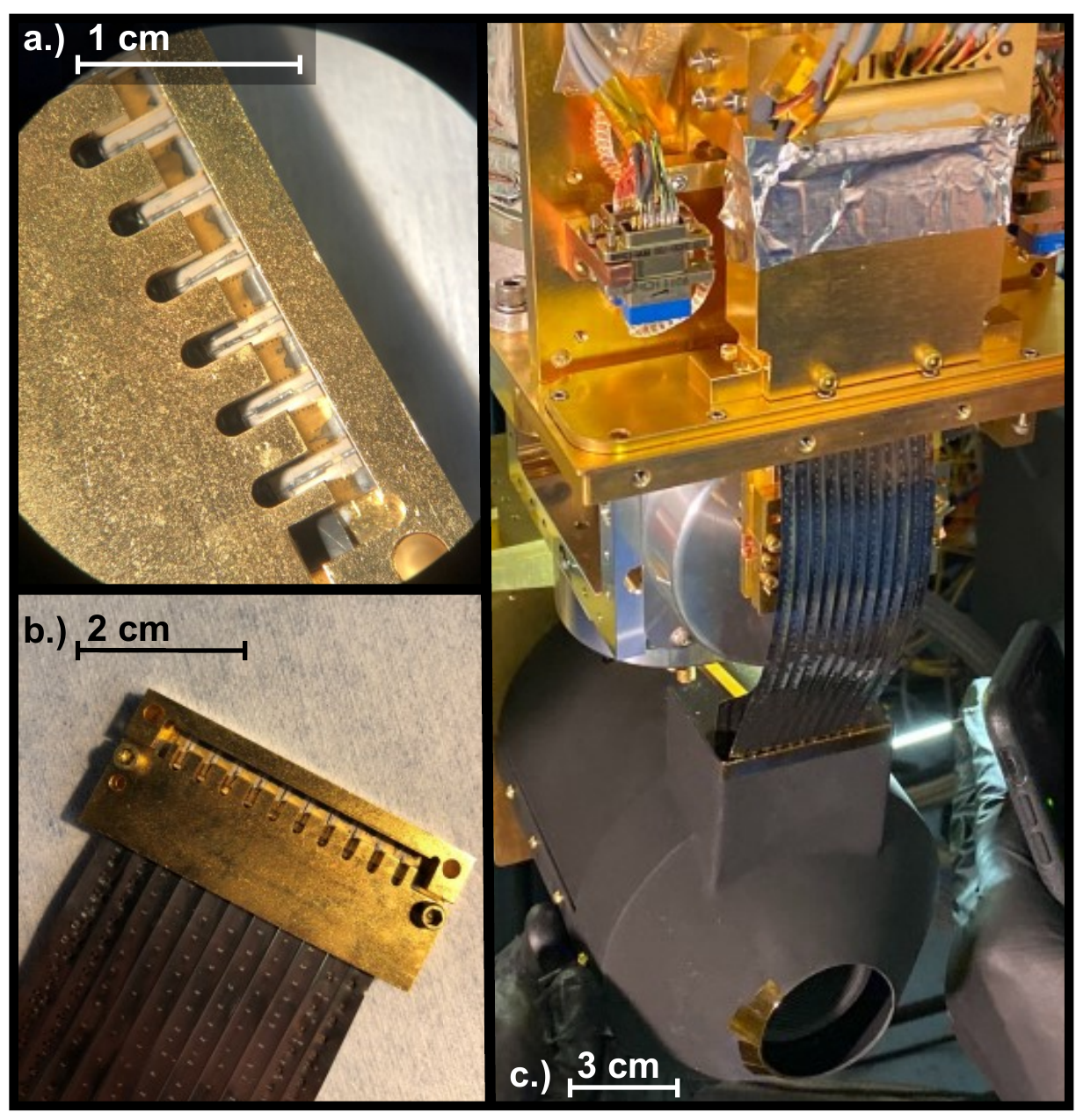}
\caption{Photographs showing a.) Close-up of cable end where NbTi center conductors connect to center trace of GCPW transition board via stainless steel capillary tubing. b.) Fully assembled cable end with protruding micro spot-welded, shared Nb47Ti ground shield. c.) Fully assembled cable spanning the 3.4 K stage and the 90 mK cold ADR stage with a thermal sink at 800 mK halfway down the length of the cable in the MKID Expolanet Camera (MEC) experiment\cite{Walter2018, Walter2020TheSCExAO}. }\label{fig:picture}
\end{figure}

An optimal solution should be made from superconducting material with a transition temperature well above 4 K to maximize transmission with an encompassing ground shield to minimize cross talk and pickup. It must have a small cross-section and be made from a low thermal conductivity material. Lastly, it should be flexible, durable, and ideally cheap and easy to manufacture. Such a structure is difficult to realize because few materials have the desired properties and often are difficult to work with and interface with connectors.

In this paper we present a superconducting FLexible coAXial ribbon cable (FLAX) which uniquely satisfies the aforementioned criteria. We developed this solution to carry broadband signals for 10 000+ pixel multiplexed Microwave Kinetic Inductance Detector (MKID) arrays for exoplanet detection operating at 90 mK \cite{Walter2020TheSCExAO, Walter2019MEC:Telescope}. We expect this technology to be especially relevant for superconducting technologies requiring high detector isolation and low thermal load.

\begin{figure}
\includegraphics[width=3.5in]{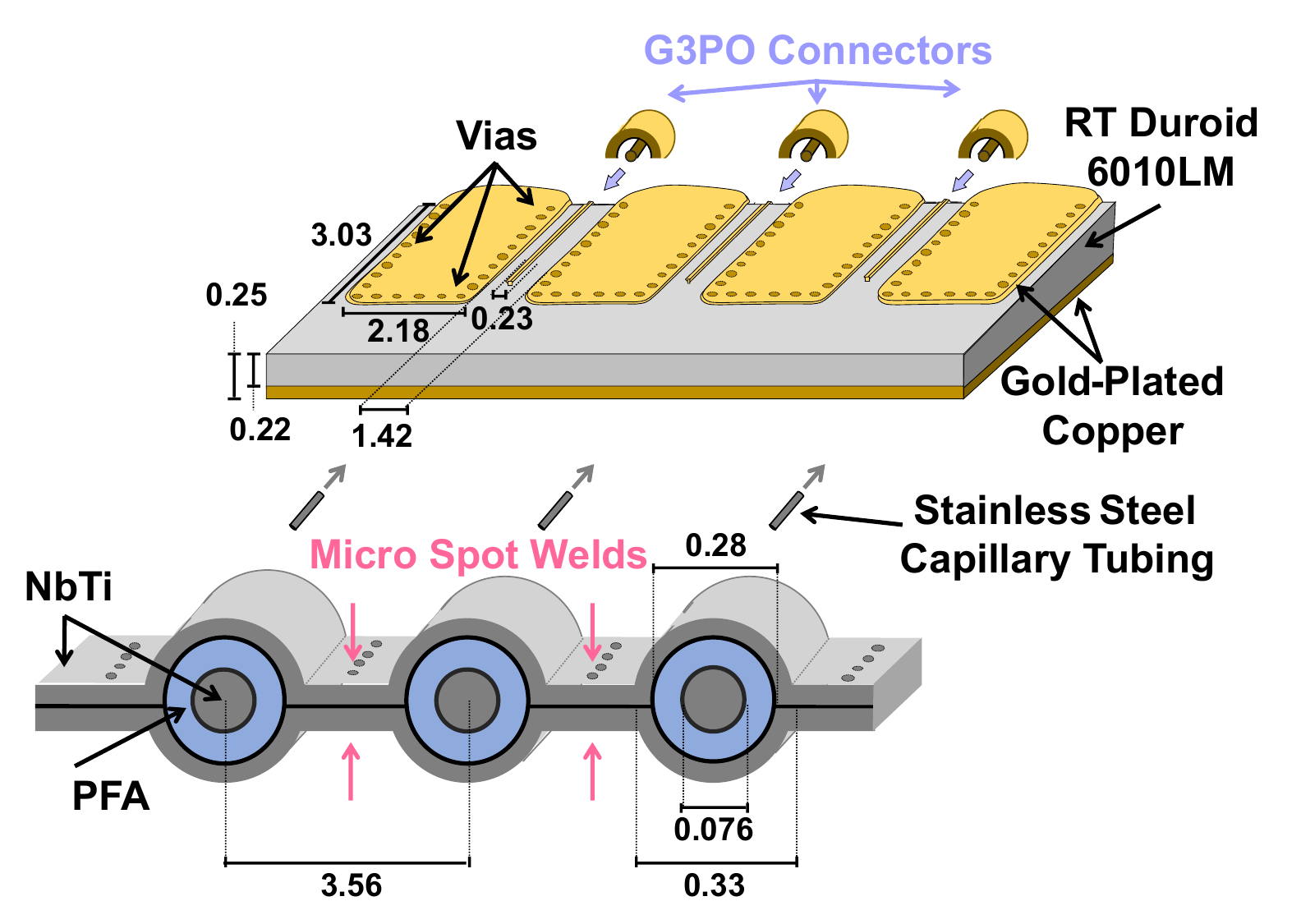}
\caption{Exploded view of cable-end assembly diagram with key dimensions shown in mm. Drawing is not to scale. From top/back to bottom/front:  G3PO half-shell connectors are soldered to the transition board. Two ground tabs with via borders and an intervening signal trace create a 50 $\Omega$ grounded coplanar waveguide. The FLAX cable center conductors are crimped into stainless steel capillary tubing and soldered to the center traces. The FLAX ground shield is spot welded to the ground tabs. The cable cross-section shows the PFA (blue) insulated NbTi (grey) wire set in semicylindrical crimps made in the shared Nb47Ti foil ground shield. The two sides of the shield are mechanically and electrically bonded with micro spot welds less than $\lambda/16\simeq$ 2 mm (at 8 GHz) apart which run in-between the traces down the length of the cable.}\label{fig:Crosssection}
\end{figure}

\section{FLAX Design and Manufacture}

 The FLAX cables are fabricated using ${\diameter}$0.076 mm [0.003"] NbTi center conductor insulated with ${\diameter}$0.28 mm [0.011"] PFA wire obtained from Supercon\footnote{Supercon Inc., 830 Boston Turnpike, Shrewsbury, MA.}. The shared outer coaxial conductor is formed with 0.025 mm [0.001"] Nb47Ti foil purchased and rolled by ATI\footnote{ATI Specialty Alloys \& Components, 1600 Old Salem Rd., Albany, OR.} and HPM\footnote{Hamilton Precision Metals, 1780 Rohrerstown Rd., Lancaster, PA.}. The wires are held in ten, ${\diameter}$0.28 mm semicylindrical crimps made 3.56 mm apart in the foil to achieve a $\sim$50 $\Omega$ characteristic impedance and 3.56 mm standard trace pitch density used by G3PO connectors available from Corning Gilbert\footnote{Corning Optical Communications, 4200 Corning Place, Charlotte, NC.} (compatible with SMP-S) (see Fig.~\ref{fig:picture},~\ref{fig:Crosssection}). The two sides of the ground shield are mechanically and electrically bonded by micro spot welds which run the length of the cable between each trace. The welds are approximately every 2 mm which is less than  $\lambda/16=$ 2.3 mm at 8 GHz (see Fig.~\ref{fig:picture},~\ref{fig:Crosssection}). 
 
 At the ends of the cable, the protruding center conductors are threaded into ${\diameter}$1.6 mm, 0.13 mm thick stainless steel capillary tubing. The tubing is crimped onto the center conductor before the assembly is soldered to the center traces of the transition board using a stainless steel soldering flux (see Fig.~\ref{fig:picture}a,~\ref{fig:Crosssection}). The transition board is a 0.25 mm thick RT/Duroid6010LM PCB with 50 $\Omega$ grounded coplanar waveguide (GCPW) geometry for increased signal isolation. Between each trace, the Nb47Ti outer conductor foil is micro spot welded to the ground tabs of the transition board while surface mount coaxial G3PO push-on connectors are soldered to the other end of the GCPW (Fig.~\ref{fig:picture}a). The cable end assembly is clamped in a 3$\times$7 cm gold-plated copper box which provides strain relief and allows for easy push-on connection of all ten traces with G3PO blind-mate bullet connectors (Fig.~\ref{fig:picture}b).

 \section{Performance Characterization}
\begin{figure}
\includegraphics[width=3.4in]{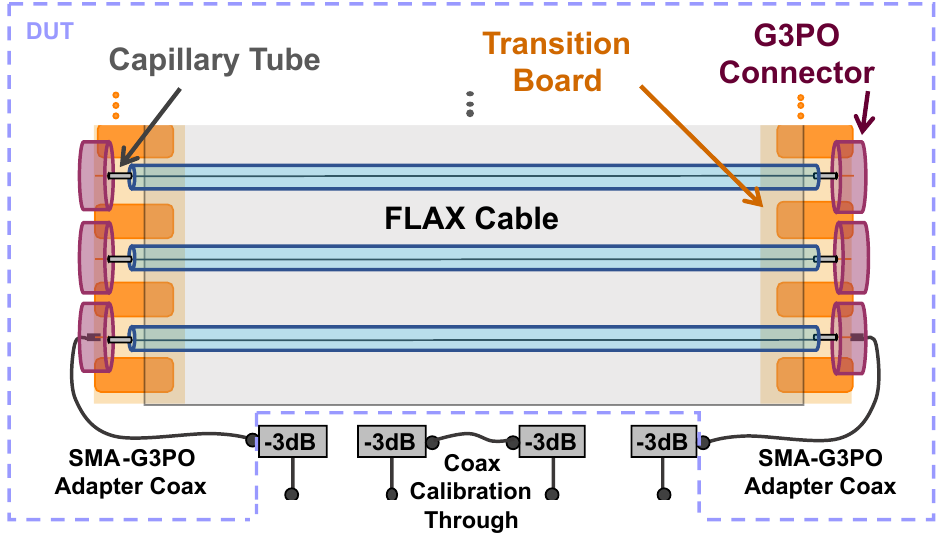}
\caption{Schematic diagram depicting FLAX attachment to the G3PO push-on connectors via a capillary tube soldered to a coplanar waveguide transition board and the device under test (DUT) circuit at 4 K. }\label{fig:DUT}
\end{figure}

Transmission loss ($S_{21}$), cross talk ($S_{41}$), and time domain reflectometry measurements were performed in a dilution refrigerator under vacuum at 4 K with a Keysight N9917A network analyzer. The device under test circuit consisted of the assembled FLAX cable with a 3 dB cryo-attenuator obtained from XMA\footnote{XMA Corporation-Omni Spectra, 7 Perimeter Road, Manchester, NH. \vskip1pt \hskip4pt P/N: 2082-6040-03-CRYO} and a 25 cm nonmagnetic SMA-to-G3PO adapter coaxial cable obtained from Koaxis\footnote{Koaxis RF Cable Assemblies, 2081 Lucon Road, Schwenksville, PA. \vskip1pt \hskip4pt P/N: AO10-CC047C-YO18} on either end (see Fig.~\ref{fig:DUT}). A Crystek\footnote{Obtained through Digikey. P/N: CCSMA18-MM-141-12} braided, semi-rigid coax  through line was used as a calibration reference. Repeated handling through the testing process revealed the cables have a minimum inside bend radius close to 2 mm and are robust to cryogenic cycling.

\subsection{Transmission}

\begin{figure*}
\includegraphics[width=\textwidth]{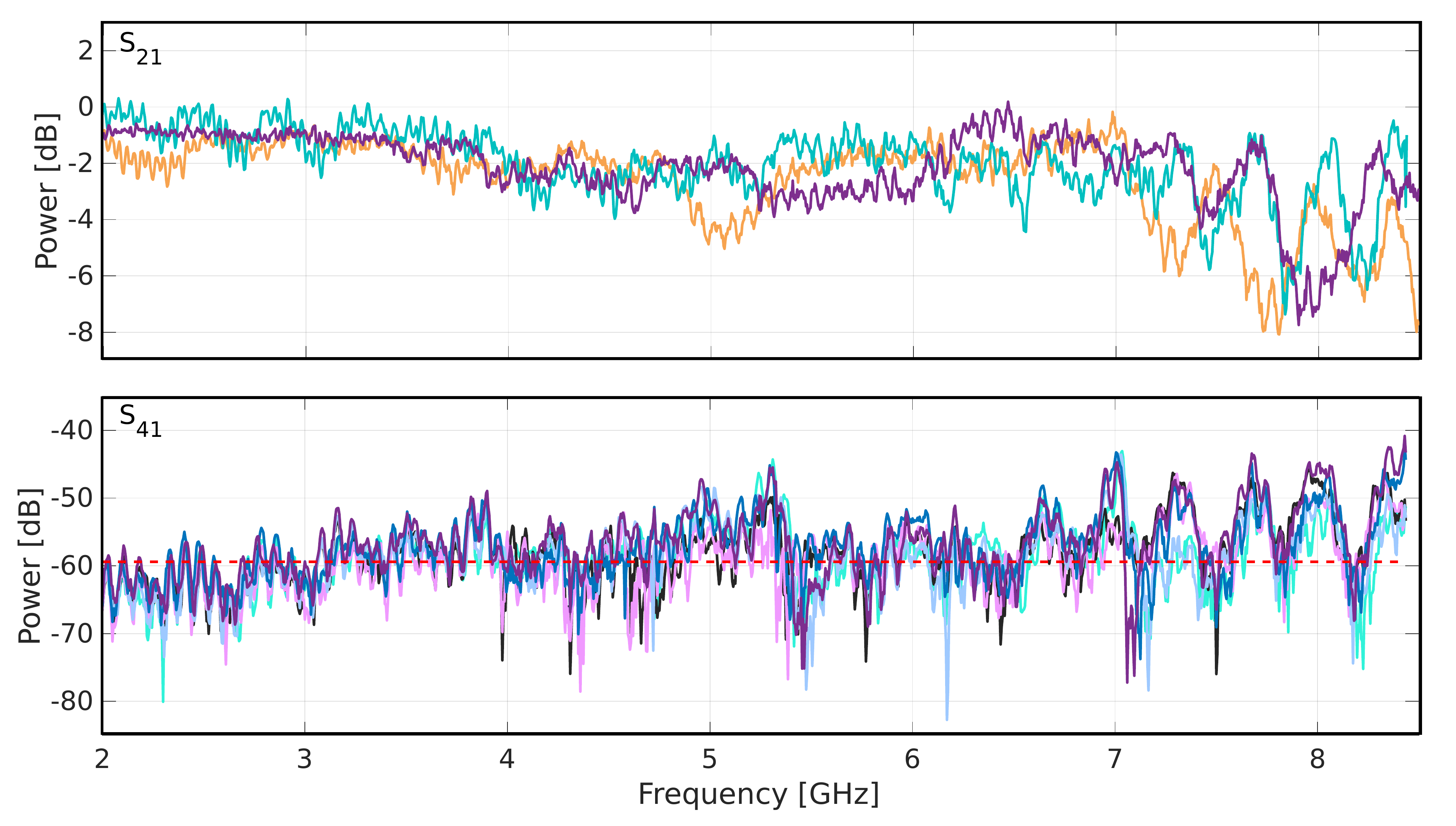}
\caption{Top: $S_{21}$ (transmission) measurement of sample FLAX traces from various cables at 4 K. Bottom: $S_{41}$ (nearest neighbor forward cross talk) measurement of sample FLAX traces from the same cable at 4 K. The average cross talk level is given by the dashed red line.}\label{fig:transmission}
\end{figure*}

\begin{figure}
\includegraphics[width=3.4in]{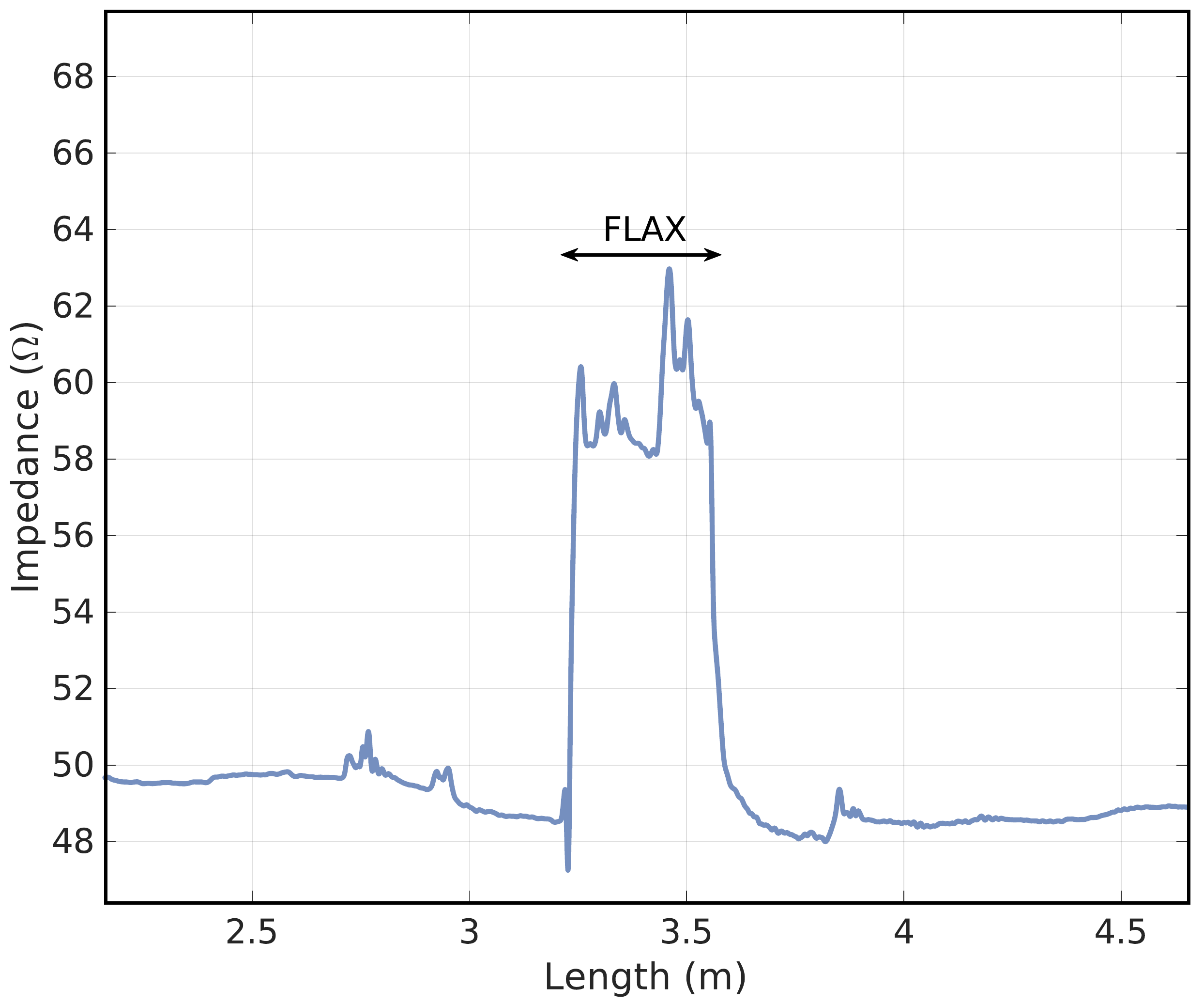}
\caption{A typical time domain reflectometry (TDR) measurement of the cryogenic signal path showing the characteristic impedance at lengths along the signal path. Commercially available 50 $\Omega$ standard coaxial cables border the FLAX cable highlighted by the double arrow. Note the TDR measurement is accurate to $\pm$ 3 $\Omega$. }\label{fig:TDR}
\end{figure}

Ripples in the FLAX transmission suggest standing wave modes are present on the traces which is indicative of an impedance mismatch between the FLAX cable and the 50 $\Omega$ circuit (see Fig.~\ref{fig:transmission}). The transmission ripples are not uniformly harmonic which suggests the impedance is changing with length along each trace. This could be explained by flaws in micro spot welding placements along the cable which determine the distance between the inner and outer coaxial conductors and therefore the characteristic impedance. The characteristic impedance of the traces were probed using a time domain reflectometry measurement adjusted for loss (see \cite{Gisin2017CharacterizingInstrument} for details on loss correction) which confirmed the impedance varies from 55--65$\pm$3 $\Omega$ along the traces (see Fig.~\ref{fig:TDR}). This mismatch at various points in the cable launches reflected waves which contribute to the observed ripple.

We hypothesize an additional factor contributing to the impedance mismatch originates in the intermediate regions of the cable ends where the center conductor exits the foil sheath and transitions onto the GCPW transition board (see Fig.~\ref{fig:picture}, a.). After exiting the ground shield, the exposed wire can act as an inductor. Previous work done by our group shows inductance on the input and output of a transmission line causes ripples which increase in magnitude at higher frequencies \cite{Walter2018}. This is because the impedance of a perfect inductor grows linearly with frequency, i.e., $Z_L = j\omega L$. With each successive cable iteration, manufacturing techniques improved, the length of exposed wire was shortened, and the frequency-dependent ripple amplitude diminished. The use of a capillary tube to pin the hair-like center conductor close to the transition board dramatically reduced the cable end inductance.

Using the peak of the ripple, we report the loss of the 30 cm cable at 8 GHz to be roughly 1 dB which is slightly higher than the 0.5 dB/m loss reported by commercially available superconducting coaxial cables \cite{CryoCoaxCryogenicCryoCoax, KEYCOMSuperconductingCables}. This difference cannot be explained by a difference in cable materials or geometry \cite{Emerson2017PTFEDifferences}. Likely, the source of our additional loss is the impedance mismatch caused by manufacturing imperfections which produce reflections in the cable and off the ends as described above.

\begin{table*}[!htbp] 
\caption{ Summary of thermal, mechanical, and microwave properties of superconducting coaxial ribbon cable, laminated microstrip cable, and best commercially available superconducting coaxial cables}\label{table:properties}
\vskip2pt
\centerline{
\vbox{\offinterlineskip
\hrule
\halign{&\vrule#&
\strut\quad#\hfil\quad\cr
&\strut&&\multispan3\hfil {\bf Thermal Load\footnote{} }\hfil&&\multispan9\hfil {\bf Mechanical}\hfil &&\multispan3\hfil {\bf Microwave}\hfil &\cr
&{\bf Cable}&&\multispan3\hfil per trace [nW] \hfil&&\multispan9\hfil All Dimensions [mm] \hfil && \multispan3\hfil Values at 8GHz \hfil &\cr
& &&\multispan3\hfil 100mK to \hfil&& Trace && OD && Min. Inside && Conductor && Dielectric && Cross Talk  \hfil &&
Attenuation\footnote{} &\cr
&\omit && 1 K & &4 K && Pitch && (${\diameter}$) && Bend Radius && Material && Material && [dB] && [dB] &\cr
height2pt&\omit&&\omit&&\omit&&\omit&&\omit&&\omit&&\omit&&\omit&\cr
\noalign{\hrule}
height2pt&\omit&&\omit&&\omit&&\omit&&\omit&&\omit&&\omit&&\omit&\cr
& FLAX && $16$ && $800$ && $3.556$ && $0.376$ && $2$ && Nb47Ti && PFA && $-60$ && $1$ &\cr
& CryoCoax && $26$ && $1400$ && $>$13 && $0.90$0 && $3.2$ && NbTi && PTFE && N/A && $<0.5$ &\cr
& KEYCOM && $34$ && $1800$ && $>$13 && $0.860$ && $8$ && NbTi && PTFE && N/A && $<0.5$ &\cr
& Nikaflex && $16$ && $460$ && $3.556$ && $0.198$\footnote{} && $6.4$ && Nb47Ti && Nikaflex\footnote{} && $-25$ && $1$ &\cr}
\hrule}}
\hskip20pt \footnotesize{$^8$ Computed using dimensions available from \cite{CryoCoaxCryogenicCryoCoax, KEYCOMSuperconductingCables, Walter2018} and assuming a cable length of 30 cm. } 
\vskip1pt \hskip20pt \footnotesize{$^9$ Estimated with ripple peak.}
\vskip1pt \hskip17pt \footnotesize{$^{10}$ For the microstrip geometry this is the total cable thickness.}
\vskip1pt \hskip17pt \footnotesize{$^{11}$ Kapton polyimide film manufactured by Dupont, see \cite{Walter2018} for details.}
\end{table*}

\subsection{Cross Talk}
We found the average nearest-neighbor forward cross talk to be -60 dB (see Fig.~\ref{fig:transmission}). This is roughly 30 dB lower than what we previously achieved using flexible laminated NbTi-on-Kapton microstrip cables \cite{Walter2018}. Since the cable's installation in the MKID Exoplanet Camera (MEC) at Subaru Observatory, this enhanced isolation has increased our pixel yield $\sim$20\% \cite{Walter2020TheSCExAO}. We suspect this large improvement is because the exposed microstrip geometry allows trace-to-trace coupling whereas the coaxial nature of the FLAX shields the center conductors thereby preventing signal corruption. In early iterations of the cable, we found infrequent or failed micro spot welds in the ground shield lead to much higher levels of cross talk. This leads us to conclude incorporating micro spot welds less than $\lambda/16$ apart between the traces reduces electromagnetic coupling.

\subsection{Thermal Conductivity}
\begin{figure}
\includegraphics[width=3.4in]{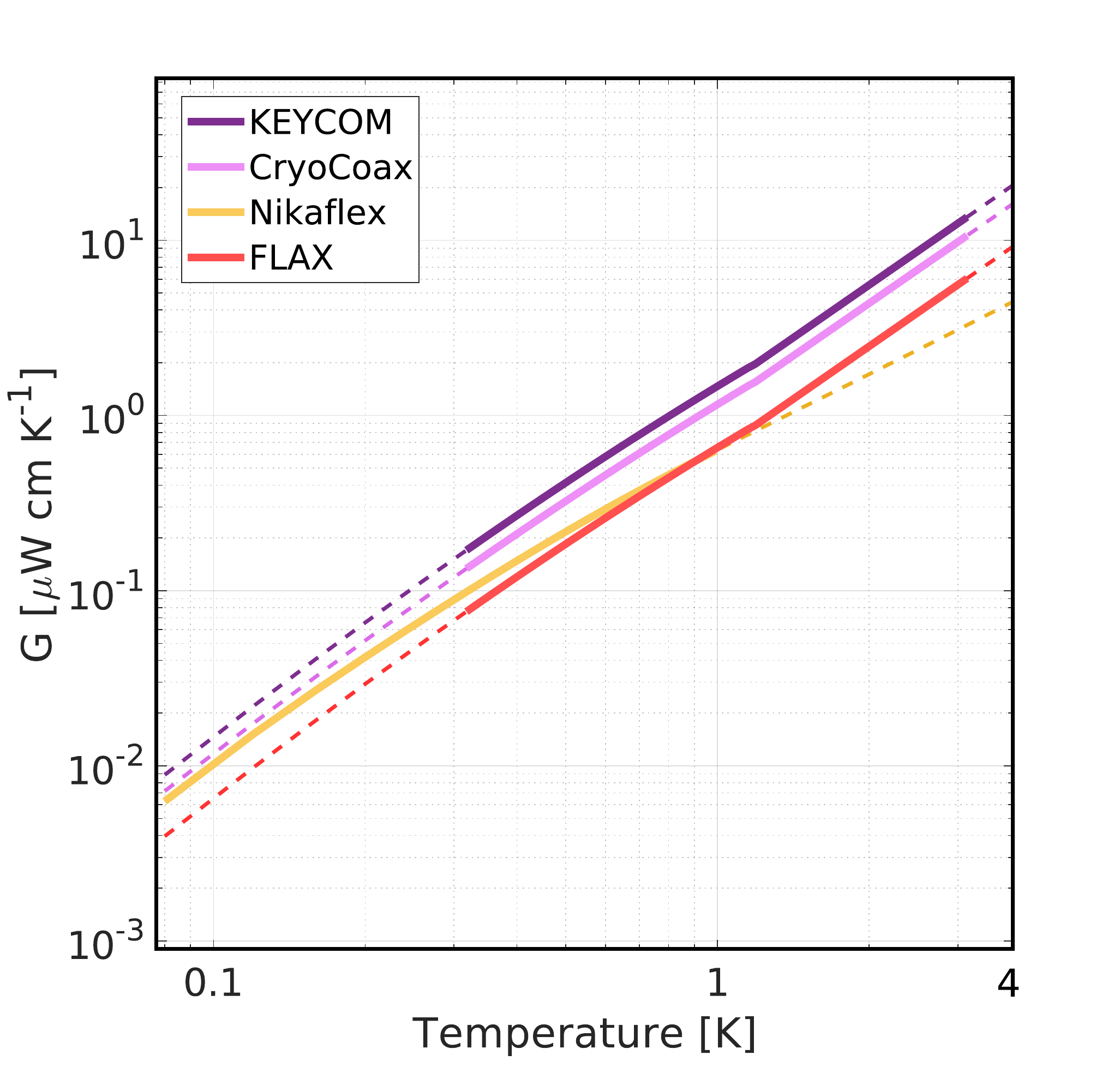}
\caption{We computed a cable thermal conductivity $G(T)$ in units of $\mu \mathrm{W cm K}^{-1}$ by summing the thermal conductivity of each constituent material weighted by the cross-section \cite{Kushino2005ThermalDetectors, Walter2018}. The cable previously developed by our lab (Nikaflex, gold) is compared with the subject of this paper (FLAX, salmon), and two commercial options by KEYCOM (P/N: NbTiNbTi034, burgundy) and Cryocoax (P/N 5139-P1NN-611-100P, pink). Solid lines are computed using literature values for Nb47Ti\cite{Daal2019PropertiesApplications, Olson1993}, PTFE\cite{Kushino2005ThermalDetectors}, Nikaflex (Kapton polyimid film)\cite{Walter2018, Kellaris2014Sub-kelvinExperiments}, and Pyralux \cite{Daal2019PropertiesApplications}. PTFE values were used to estimate the PFA dielectric in the FLAX cable\cite{Emerson2017PTFEDifferences}. Dashed lines indicate extrapolation.}\label{fig:heatload}
\end{figure}

Following previous convention, a cable thermal conductivity, $G(T)$, was computed by summing literature values of constituent materials weighted by their cross-sections (see Fig.~\ref{fig:heatload}) \cite{Kushino2005ThermalDetectors,Walter2018}. We compare our superconducting coaxial ribbon cable to two commercially available superconducting coaxial cables as well as our lab's previously developed laminated NbTi-on-Kapton microstrip cables \cite{Walter2018}. We estimate the thermal conductivity of the PFA dielectric present in the flexible coaxial ribbon cables using PTFE; the same dielectric used in the two commercial solutions \cite{Emerson2017PTFEDifferences}. The smallest commercially available superconducting coaxial cables from KEYCOM\footnote{KEYCOM Corp. 3-40-2 Minamiotsuka,Toshima-ku Tokyo. \vskip1pt \hskip5pt P/N: NbTiNbTi034} and CryoCoax\footnote{CryoCoax - Intelliconnect, 448 Old Lantana Road, Crossville, TN. \vskip1pt \hskip5ptP/N: 5139-P1NN-611-100P} were chosen for comparison. The electrical and thermal properties of the cables are summarized in table~\ref{table:properties}. 

The heat load from one temperature stage to another can be computed by integrating values in Fig.~\ref{fig:heatload} from $T_1$ to $T_2$ ($T_1<T_2$) and dividing by the cable length. The ten-trace FLAX cables are currently installed in the MEC experiment where they span 33 cm from the 3.4 K stage to the 90 mK cold ADR stage with a thermal sink at 800 mK about halfway down the length of the cable \cite{Walter2019MEC:Telescope}. We estimate they generate a thermal load of $\sim$200 nW on the 90 mK cold ADR stage. This is about equivalent to the thermal load created by the Nikaflex cables and approximately half the computed heat load of either commercial option.


\section{Conclusion}
We have manufactured a superconducting flexible coaxial cable capable of delivering microwave signals between temperature stages with minimal loss, cross talk, and heat conduction. Strong signal isolation is especially important for our application of moving 4-8GHz servicing 10 000+ multiplexed sensors across temperature stages. The FLAX cable represents a 30 dB improvement in cross talk as compared to our group's previously developed NbTi-on-Kapton microstrip cables. This enhanced isolation facilitated a $\sim$20\% increase in MKID pixel yield in the MEC experiment\cite{Walter2020TheSCExAO}. We expect these results will be especially useful for high-density microwave superconducting detector arrays requiring strong signal isolation.

The cable technology presented in this paper also has very low thermal conductivity. For a given thermal budget, the FLAX cables allow for twice as many detectors as the leading commercial option. The reduced heat load combined with the push-on, small form factor connectors and reduced trace pitch allow for increased detector density in a cryogenic system.

We found an attenuation of $1$ dB at 8 GHz with $\sim$3 dB ripples which is at worst 2x more loss than commercial options. This magnitude of ripples and loss do not impact our array on the input side as we can drive microwave resonators (MKIDs) located at transmission dips with higher power than their frequency neighbors. However, these features degrade the overall signal to noise ratio on the output. Ripples and loss may become prohibitive for systems operating at frequencies over 8 GHz or systems constrained by amplifier dynamic range. Insertion loss and ripples can be reduced by improving manufacturing precision in the forming of the NbTi foil crimps and location of micro spot welds. Alternative methods to join the push on connectors and traces, e.g., brush plating the NbTi center conductor with an easily solderable material such as nickel may also improve the impedance match.

Lastly, we note these cables are relatively easy to fabricate. Many components, most notably the fine, NbTi center conductor wire, are commercially available. All cable iterations were manufactured in-house at the University of California, Santa Barbara. Ten trace FLAX can be assembled in two days. Overall, we find this cable technology to be superior to commercial options for our applications building high-density superconducting detector arrays.

\section*{Acknowledgment}
J. P. Smith is supported by a NASA Space Technology Research Fellowship under grant number 80NSSC19K1126.
\bibliographystyle{IEEEtran}
\bibliography{Flax_Paper}
\end{document}